\documentstyle[12pt,pra,aps]{revtex}
\begin{document}

\title{Reply to Comment of  Gazdzicki and Heinz 
on Strangeness Enhancement in $p+A$ and $S+A$ \footnote{
This work was supported in part by the Director, Office of Energy
Research,Division of Nuclear Physics of the Office
of High Energy and Nuclear Physics of the U.S Department of 
Energy under Contract No. DE-FG02-93ER40764, DE-AC03-76SF00098}}
\vspace{2ex}
\author{M. Gyulassy$^1$, V. Topor Pop$^{1,2}$,and X. N. Wang$^3$}
\address{\normalsize $^1$ Department of Physics,
Columbia University, New York, N.Y. 10027\\
$^2$ Institute
 for Space Sciences,P.O.Box MG - 6,Bucharest, Romania\\
$^3$ Nuclear Science Division,Lawrence Berkeley Laboratory,\\
University of California, Berkeley, CA 94720}
\maketitle
\vspace{0.5in}
\begin{abstract}
The  Comment of Gazdzicki and Heinz\cite{comment}
is flawed because their assumed baryon stopping
power in $pA$ is inconsistent with data and because
they ignored half the analysis based on the VENUS model.
The Comment continues the misleading  
presentation of strangeness enhancement by focusing
on ratios of integrated yields. Those ratios discard essential experimental
information on the rapidity dependence of  produced
$\Lambda$ and obscure  discrepancies between
different data sets. Our conclusion remains
that the NA35 minimum bias data on $p+S\rightarrow\Lambda +X $ indicate
an anomalous enhancement of central rapidity 
strangeness in few nucleon reactions
that points to non-equilibrium dynamics as responsible
for strangeness enhancement in nuclear reactions.
\end{abstract}

\vspace{2ex}
The Comment\cite{comment} of  Gazdzicki and Heinz
 addresses  our recent analysis\cite{topor}, where we
concluded that if the NA35 data\cite{na35} on $p+S$ and $S+S$ are correct,
then strangeness enhancement must be due
to new non-equilibrium multiparticle production dynamics.
We further concluded that it is 
misleading  to analyze strangeness enhancement in terms
of  ratios of integrated yields such as
$E_s=\langle \Lambda + K + \bar{K}\rangle/\langle \pi
\rangle$ as advocated in Refs.\cite{na35,gazdzicki}.
Such ratios discard essential experimental information
on the rapidity distribution of produced strange particles which 
contradict the conclusions drawn from those ratios alone.
Such ratios also obscure 
striking discrepancies between different experiments
which indicate that additional experiments, especially with
$pA$,  will be needed to 
understand the strangeness enhancement phenomenon.

The focus of our work was the
anomalous rapidity distribution of 
$\Lambda$ reported by NA35\cite{na35}
for minimum bias $p+S$ reactions. Our analysis  used two
different microscopic models, HIJING\cite{hijing} and VENUS\cite{venus},
to quantify the differences between $p+p$, $p+A$ and $A+A$ reactions.
Gazdzicki and Heinz \cite{comment} missed entirely
the VENUS half of our analysis
and argued incorrectly and without calculation
on the role of  baryon stopping power to explain the $pA$ data.

In this reply, we contradict two key
aspects of the argument presented in \cite{comment}: 
(1) their assertion that there is no strangeness enhancement in $p+S$ reactions
and (2) their assertion that our analysis  \cite{topor} 
should be dismissed  because one of the  models
used, HIJING, underestimates the  baryon stopping power of nuclei
and overestimates the  rapidity density of $\Lambda$'s
in the fragmentation regions of $p+p$. 
Other points raised 
in \cite{comment} in connection with error bars on integrated
yields  and the $p_T$ acceptance cuts are irrelevant
to our conclusions and  will not be addressed here.

First we review the  NA35 data on minimum bias $p+S$ \cite{na35} 
that contradict assertion (1). Then we review the
known baryon stopping power from  $p+A\rightarrow p+X$\cite{busza}
and discuss the significance of the VENUS model simulations
to contradict assertion (2). 

The primary motivation for our work \cite{topor} was 
 the very unusual
rapidity distribution of $\Lambda$'s reported by NA35\cite{na35} for the {\em
minimum bias} $p+S$ reactions.
 The central $(y\approx 3$) rapidity density $(0.06\pm
0.01)$ of $\Lambda$'s produced in $p+S$
was reported to be four times more than that ($0.016\pm 0.0005$) in 
$p+p$\cite{pp}
 (Fig 1a,b
\cite{topor}). This is very surprising since the  cross section for the
selected $p+S$  ($n_{ch}>5)$ events was 470 mb,  
which  corresponds to a  minimum bias
and not central trigger. Taking the diffuse nuclear surface into account 
that trigger only constrains events to have  impact parameters less than
5 fm.  For this weak trigger, the mean number of interacting $S$ nucleons
is only $\nu \approx 2.2 $.
The $p+S$ reaction therefore only tests the difference between multiparticle
production dynamics in two nucleon
$(\nu=1)$ and few  nucleon $(\nu=2-3)$ reactions. 

>From extensive $p+A\rightarrow p+X$
systematics\cite{busza} the average baryon rapidity shift in 
$p+A$ reactions grows slowly as $\Delta y_B\approx 1+(\nu-1)/3$.
In $p+S$ therefore the leading baryon rapidity shift is only
a half a unit greater than in $p+p$.
The qualitative discussion on baryon stopping in \cite{comment}
is wrong. Stopping power cannot account for the
strong suppression of $\Lambda$ production for $y>4.5$ in $p+S$.
Also the number of collisions $\nu$ is too small to
account for the factor of four enhancement of the central rapidity
density. In the target fragmentation region $(y\approx 1)$ there
is also a factor of four enhancement of the $\Lambda$ density
in $pS$ relative to $pp$.
If the $p+S$ data are correct, then  strangeness
enhancement already occurs in few nucleon processes and therefore
must be due to new non-equilibrium dynamics.

In stating the conclusion,  we carefully pointed
out however the fact that the NA35 data on $p+S$ differ substantially from
earlier  data on $pAr$ by NA5\cite{na5} and preliminary 
NA36\cite{na36} $pPb$ data.
The earlier
NA5 analysis \cite{na5} showed  that both the Dual Parton and Lund models
could account easily for the factor of two enhancement
of the central $\Lambda$ rapidity densities in
$p+Ar$ and similar enhancement in $p+Xe$ and $\bar{p}+Xe$.
However the NA5 $p+Ar\rightarrow \Lambda+X$ central density is a factor of 
two lower than found in $p+S$ by NA35. 
The preliminary NA36 data also
suggest that the mid-rapidity $\Lambda$ density in  $p+Pb$ may be a factor of
two lower than in $p+S$ of NA35. 
Furthermore, the $p+A\rightarrow\Lambda+X$ systematics of
Ref.\cite{brick} show that the enhancement of
$\Lambda$ production in $p+A$ increases linearly with
the number of secondary collisions. Therefore, strangeness enhancement
builds up gradually in $pA$ according to those other data, and is
anomalously enhanced in the central region according to NA35 data.
The ratios of integrated yields
however hides these discrepancies completely, and that alone should
be  enough 
to disqualify such ratios as useful observables to
study strangeness production in nuclear reactions.

In \cite{topor} we  assumed the validity of the
NA35 $p+S$ data since otherwise the validity of the even 
more spectacular central
$S+S$ data would also have to be questioned (the NA35,NA36 discrepancy
in central $SS$ also remains unresolved). If, on the other hand,
 data on strangeness production eventually converge toward the 
lower NA5, NA36 range, then indeed our conclusions would have to
change to a more pessimistic one - namely - conventional hadron cascade models 
could then account for most of the strangeness production (see e.g. Fig 13-15
\cite{na36}). In that case  there would of course 
be no  need to debate further this topic.

Finally, we contradict assertion (2)
of the comment\cite{comment}. 
The overestimate  of the fragmentation region $\Lambda$ rapidity
density with HIJING in  $p+p$ 
in spite of the  accurate agreement with the mid-rapidity
density  is used   to dismiss our model analysis.
However, the fact that unlike in $p+p$ both HIJING and VENUS  
fit the target fragmentation $(y=1)$ 
peak $p+S$ in Fig 1b\cite{topor} is itself direct
evidence for strangeness enhancement. We note that the enhancement of
 target fragmentation region 
is consistent with NA5 \cite{na5} and Ref.\cite{brick} and is not 
controversial.
However, the strong suppression
of $\Lambda$ in the projectile fragmentation region
($y>4.5$) in $p+S$ is not found in either
model.  The authors of \cite{comment}
missed entirely the significance of the VENUS model calculation
which reproduces well the  $p+A\rightarrow p+X$
stopping power measurements\cite{venus}.
In fact VENUS reproduces well the NA35 central $SS\rightarrow pX$
distribution.
 HIJING is too strongly peaked about
the mean rapidity loss while VENUS distributes baryons more
broadly about that mean in accord with data.
The small shift of the VENUS curves in the fragmentation
regions (Fig. 1b\cite{topor}) relative to HIJING show
that correct baryon stopping cannot account for the
strong suppression of $\Lambda$'s with $y>4.5$ in $p+S$.
These calculations prove that the qualitative
 arguments used by the authors
to account for the enhanced central rapidity $\Lambda$'s
three units of rapidity away from the fragmentation regions
due to baryon stopping are wrong.
Furthermore, as we emphasized in \cite{topor} the agreement
of the central rapidity density with $pS$ data by VENUS is itself another
 independent
proof that the central rapidity density   in that reaction 
is  anomalous
since VENUS overestimates the $p+p\rightarrow \Lambda+X$ central 
density by a factor of two.

The analysis of ratios of integrated yields based on simplistic fireball models
and concepts \cite{gazdzicki,rafelski} is misleading since the
glaring discrepancies between the assumed differential distributions and data
(and between data sets) can be well hidden from scrutiny. There can be no
justification for throwing away valuable 
experimental information on differential
yields. Conclusions about the necessity of quark-gluon plasma production based
on production ratio systematics alone are thus not well founded.  We have shown
in \cite{topor} for example that at least one dynamical model (VENUS) can
account for the anomalous central $SS$ Lambda production distributions.
Contrary to the assertion in \cite{comment} strangeness enhancement does occur
in $pA$ at least in the target fragmentation region\cite{brick} and possibly in
the central region as well\cite{na35}.  However, better data on $pA$ will
be required to resolve experimental discrepancies
and to understand the non-equilibrium dynamical processes responsible
for that enhancement.  The search for unambiguous signatures of a new states of
matter requires untangling complex and as yet poorly understood multiparticle
dynamical effects that can forge naive plasma signatures.  Only the detailed
systematics and correlations between {\em differential} observables will be
useful in that search.

\end{document}